\pdfoutput=1 

\documentclass[11pt]{article}
\usepackage{moriond}

\def\Journal#1#2#3#4{{#1} {\bf #2}, #3 (#4)}

\def\be{\begin{equation}}
\def\ee{\end{equation}}
\def\bea{\begin{eqnarray}}
\def\eea{\end{eqnarray}}

\begin{document}
\vspace*{4cm}
\title{CHARTING THE TEV MILKY WAY: H.E.S.S. GALACTIC PLANE SURVEY
  MAPS, CATALOG AND SOURCE POPULATIONS} 

\author{ S. CARRIGAN$^{1}$, F. BRUN$^{1}$, R.C.G. CHAVES$^{2,1}$,
  C. DEIL$^{1}$, H. GAST$^{3,1}$ AND V. MARANDON$^{1}$  FOR THE
  H.E.S.S. COLLABORATION} 

\address{ 
  $^{1}$Max-Planck-Institut f\"ur Kernphysik, P.O. Box 103980, D 69029
  Heidelberg, Germany \\ $^{2}$DSM/Irfu, CEA Saclay, F-91191
  Gif-Sur-Yvette Cedex, France \\  $^{3}$I. Physikalisches Institut B,
  RWTH Aachen University 
}

\maketitle
\abstracts{
Very-high-energy (VHE, E$>$100 GeV) $\gamma$-rays provide a
unique view of the non-thermal universe, tracing the most violent
and energetic phenomena at work inside our Galaxy and beyond. 
The latest results of the H.E.S.S. Galactic Plane 
Survey (HGPS) undertaken by the High Energy Stereoscopic System
(H.E.S.S.), an array of four imaging atmospheric Cherenkov telescopes
located in Namibia, are described here. The HGPS aims at the detection of
cosmic accelerators with environments suitable for the production of
photons at the highest energies and has led to the discovery of an
unexpectedly large and diverse population of over 60 sources of TeV
gamma rays within its current range of l = 250 to 65 degrees in
longitude and $|$b$|<$ 3.5 degrees in latitude. The data set of the HGPS
comprises 2800 hours of high-quality data, taken in the years 2004 to
2013. The sensitivity for the detection of point-like sources, assuming
a power-law spectrum with a spectral index of 2.3 at a
statistical significance of 5 $\sigma$, is now at the level of 2\%
Crab or better 
in the core HGPS region. The latest maps of the inner Galaxy at TeV
energies are shown alongside an introduction to the first
H.E.S.S. Galactic Plane Survey catalog. Finally, in addition to an
overview of the H.E.S.S. Galactic source population a few remarkable,
recently discovered sources will be highlighted.
}

\section{Introduction}
The superior sensitivity of the latest generation of imaging atmospheric 
Cherenkov telescopes such as H.E.S.S.~\cite{Hinton2004}, 
MAGIC~\cite{Aleksic2012}, and 
VERITAS~\cite{Weekes2002}, has resulted in the discovery of numerous 
VHE $\gamma$-ray sources, and currently more than 140 sources are 
listed in the online TeV  $\gamma$-ray catalogue 
TeVCat~\footnote{http://tevcat.uchicago.edu/}. 
The High Energy Stereoscopic System (H.E.S.S.) in particular is
ideally suited for undertaking a deep survey of our Galaxy, due to 
its high sensitivity, comparatively large field-of-view of 5$^\circ$, 
and its angular resolution of  $\sim$0.1$^\circ$. Its location in the 
Khomas highlands of Namibia allows it 
a prime view of the inner Galaxy. Here we report on the
status and latest results of the H.E.S.S. Galactic Plane Survey
(HGPS), the deepest and most 
comprehensive survey of the inner Galaxy undertaken in VHE $\gamma$-rays so
far. The latest maps are shown, alongside 
advanced methods for the suppression of cosmic-ray induced
background. The 
construction of a software framework to detect and model sources of VHE 
$\gamma$-rays in the survey data set is introduced. 
The VHE $\gamma$-ray source population in our Galaxy is dominated by
objects that are linked to the final stages in stellar evolution,
namely pulsar wind nebulae (PWNe) and supernova remnants (SNRs). For
nearly a third of the sources, however, no plausible counterpart at
other energies has been found yet, or the physical origin of the
detected emission remains unclear. A snap shot of our current
understanding of the H.E.S.S. source 
population on the Galactic plane is shown, as well as a few interesting 
examples of recent sources.

\section{Maps}

 In the HGPS, the inner Galaxy has been
systematically raster scanned using observation positions with overlapping
fields-of-view, with the main goal of discovering new VHE $\gamma$-ray
sources and enabling population studies of Galactic source classes as
a consequence. Advanced analysis techniques for background
suppression~\cite{Ohm2009,Naurois2009,Fiasson2010,Becherini2011} play a very
important role in the data analysis. After calibration and quality
selection, a multi-variate analysis technique~\cite{Ohm2009} based on
extensive air shower and image shape parameters is used to
discriminate $\gamma$-ray-like events from cosmic-ray-induced
showers. A minimum image amplitude of 160 photoelectrons is required. 

\begin{figure}
\begin{minipage}{0.5\linewidth}
\centerline{\includegraphics[width=\linewidth]{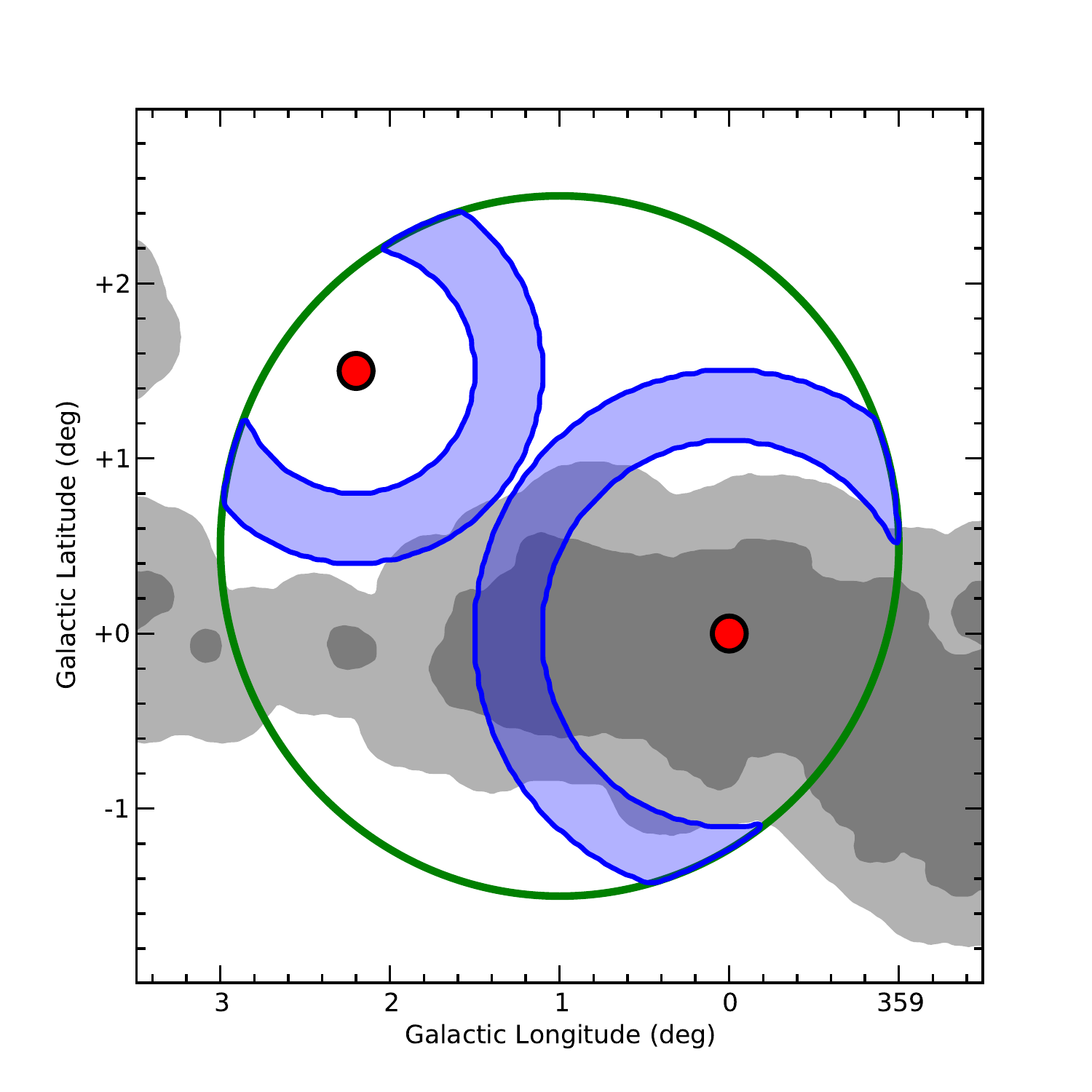}}
\end{minipage}
\begin{minipage}{0.5\linewidth}
\centerline{\includegraphics[width=\linewidth]{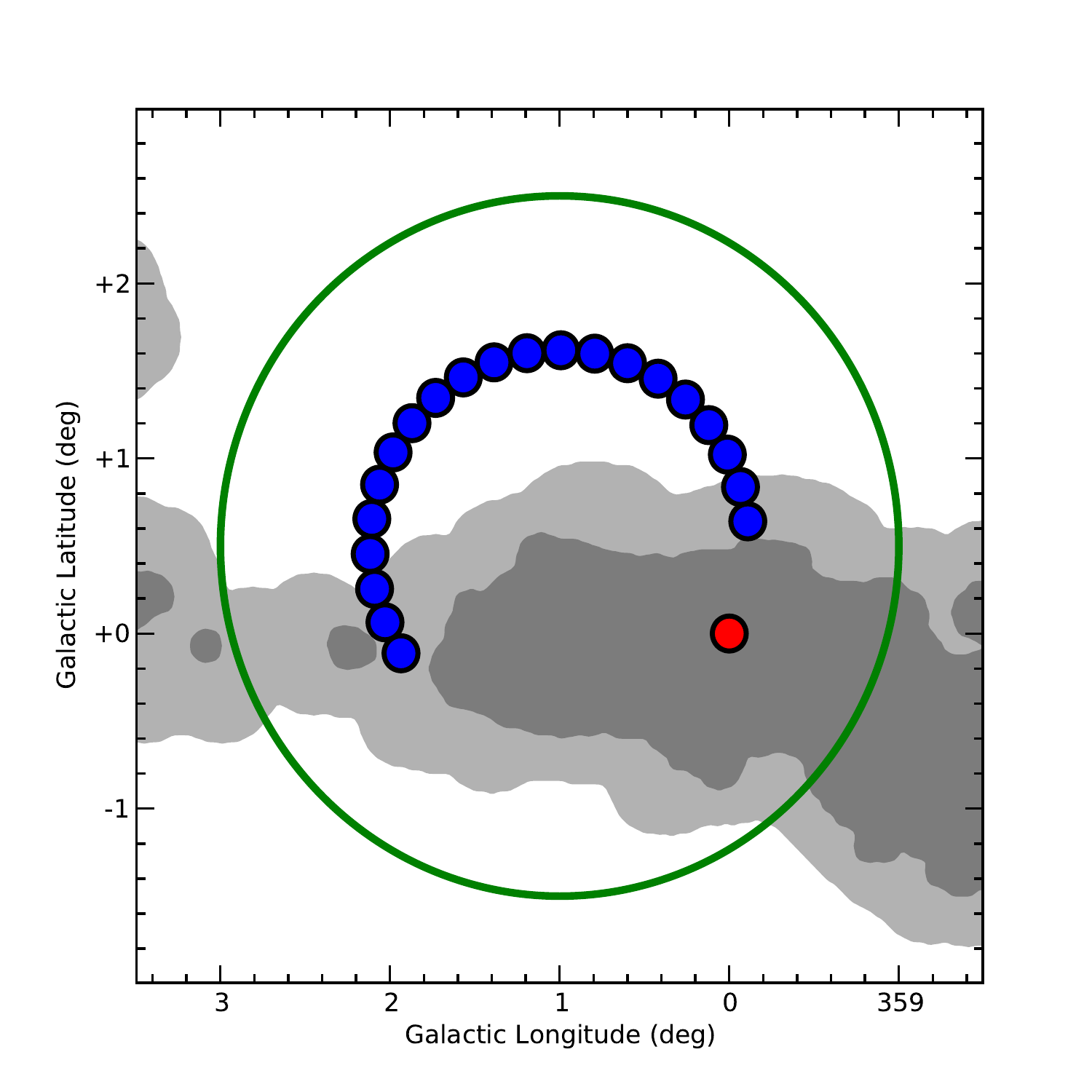}}
\end{minipage}
\caption
{Illustration of the different background
estimation methods for image and spectral analysis, as well as the
challenges the high density of extended 
sources in the inner Galaxy poses. Exclusion regions
(see Section 2 for details) are 
shown as grey areas. The field-of-view of 2$^\circ$ radius is
illustrated as green solid circles. {\it Left
  panel}: The \emph{adaptive ring background technique}.  {\it Right
  panel}: The \emph{reflected region background technique}.
}
\label{fig:excl_bg}
\end{figure}

\begin{figure}[p]
\includegraphics[width=\linewidth]{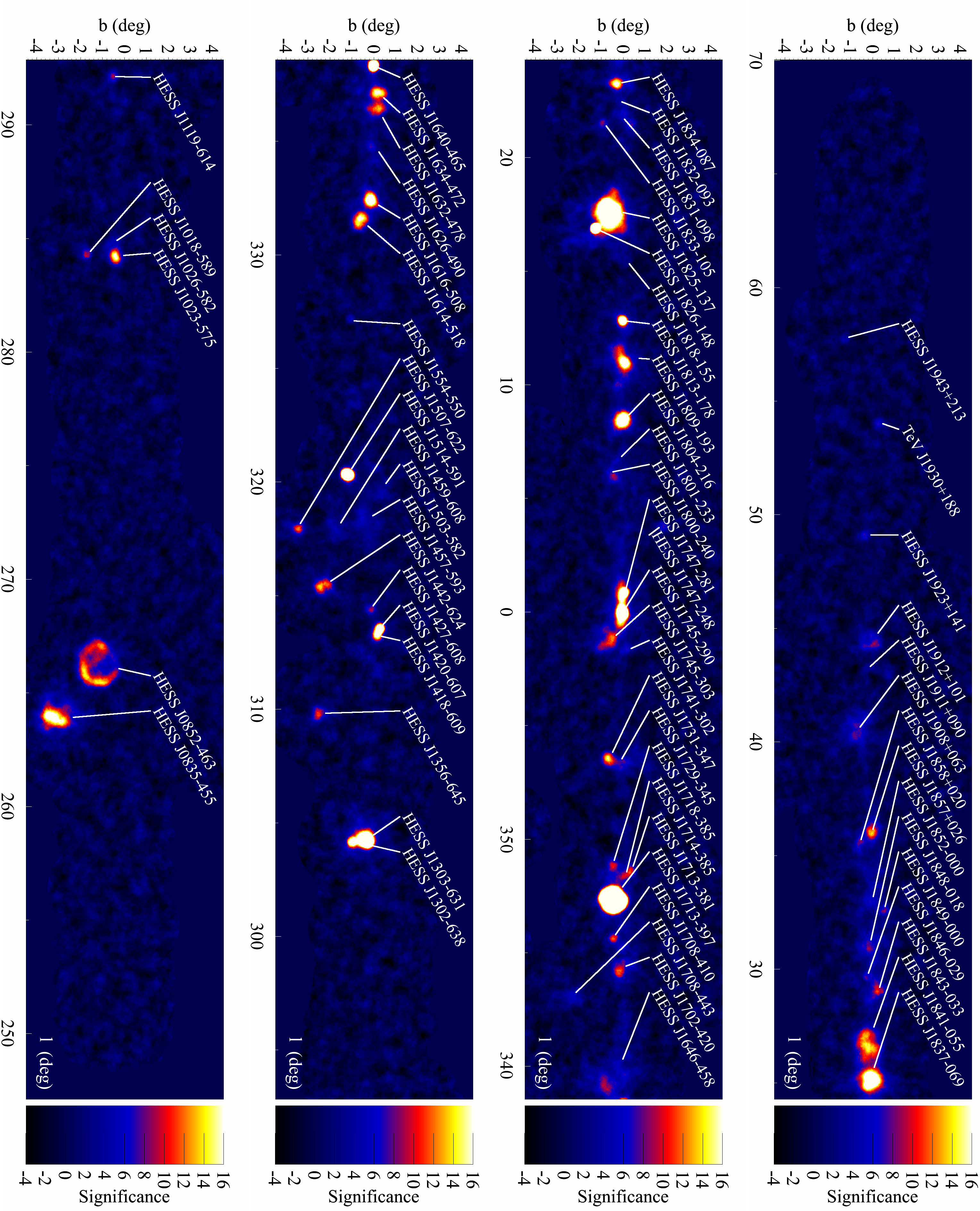}
\caption
{Significance map for the H.E.S.S.~Galactic Plane Survey. The
pre-trials significance for a correlation radius of $0.1$ deg is
shown. The colour transition from blue to red corresponds to
$\sim\!5\sigma$ post-trials 
significance. The significance has been calculated for regions on the
sky where the 
sensitivity of H.E.S.S.~for point sources ($5\sigma$ pre-trials, and assuming
the spectral shape of a power law with index 2.3) is better than
10$\,\%$ Crab. Identifiers for sources that have been described in
publications or announced at conferences are included.
}
\label{fig:significancemap}
\end{figure}

To generate maps, the remaining
background is estimated locally by the {\it ring background
technique}~\cite{Berge2007}, where for each trial source position (red
filled circles) in the field-of-view (of 2$^\circ$ radius, green circle) the
background is estimated from a ring centered on this position 
(blue shaded circles), as shown in the left panel of
Figure~\ref{fig:excl_bg}. Regions on 
the sky containing known VHE $\gamma$-ray sources (grey areas) are excluded
from background estimation. These exclusion regions are automatically
generated from significance 
maps which are smoothed with a top-hat function with radii of
0.1$^\circ$, 0.2$^\circ$ and 
0.4$^\circ$, by thresholding them at the level of 5$\sigma$ and 
dilating each excluded pixel by a further 0.1$^\circ$ ({\it
  small} exclusion regions, dark grey) or 0.3$^\circ$ ({\it large} exclusion
regions, light grey). For the {\it small} exclusion regions, the resulting
maps for 0.1$^\circ$ and 0.2$^\circ$ radius are added, excluding all
significant emission from the background but cutting quite close to
the edges of sources. For the {\it large} exclusion regions, the
resulting maps for 0.1$^\circ$, 0.2$^\circ$ and 0.4$^\circ$ radius are
added, cutting away more of any emission that is possibly extending into
the area for background estimation. The resulting {\it large}
exclusion regions, used for the production of 
maps, turn out to cover areas of the
sky that are comparatively large on the scale of the size of the
field-of-view. Therefore, as illustrated in the left panel of 
Figure~\ref{fig:excl_bg}, the ring radius is adaptively enlarged when a large 
fraction of the ring area overlaps with an excluded region, until an
appropriate ring of the same thickness is reached. The statistically
significant value for each position is then
calculated~\cite{LiMa1983}, by summing the candidate events within a
fixed and predefined correlation radius, e.g. 0.1$^\circ$ (suitable 
for point-like sources), and comparing to the estimated background level
at that position. Figure~\ref{fig:significancemap} shows the latest
significance map obtained for the survey region. 

\begin{figure}
\includegraphics[width=\linewidth]{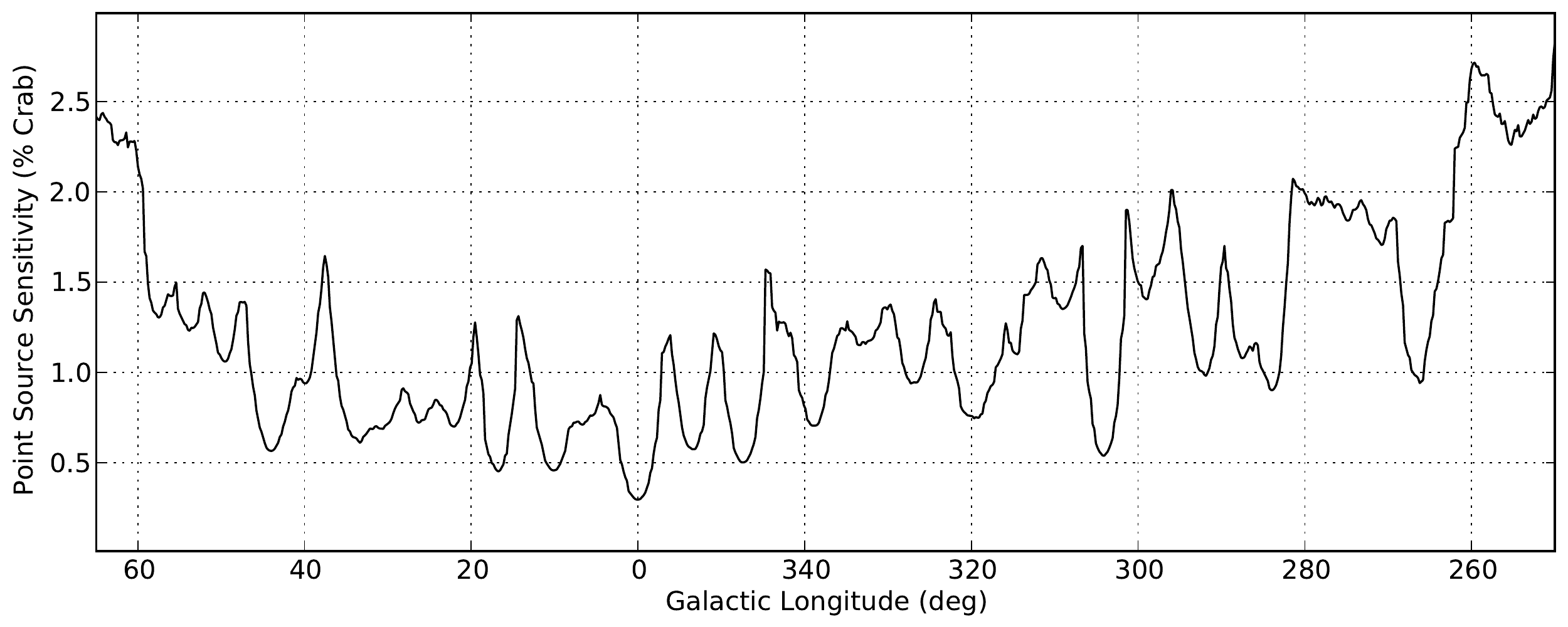}
\caption
{Profile of the sensitivity map (refer to Figure~\ref{fig:sensitivitymap})
for b=$-$0.3$^\circ$.
}
\label{fig:sensitivityslice}
\end{figure}

\begin{figure}[p]
\includegraphics[width=\linewidth]{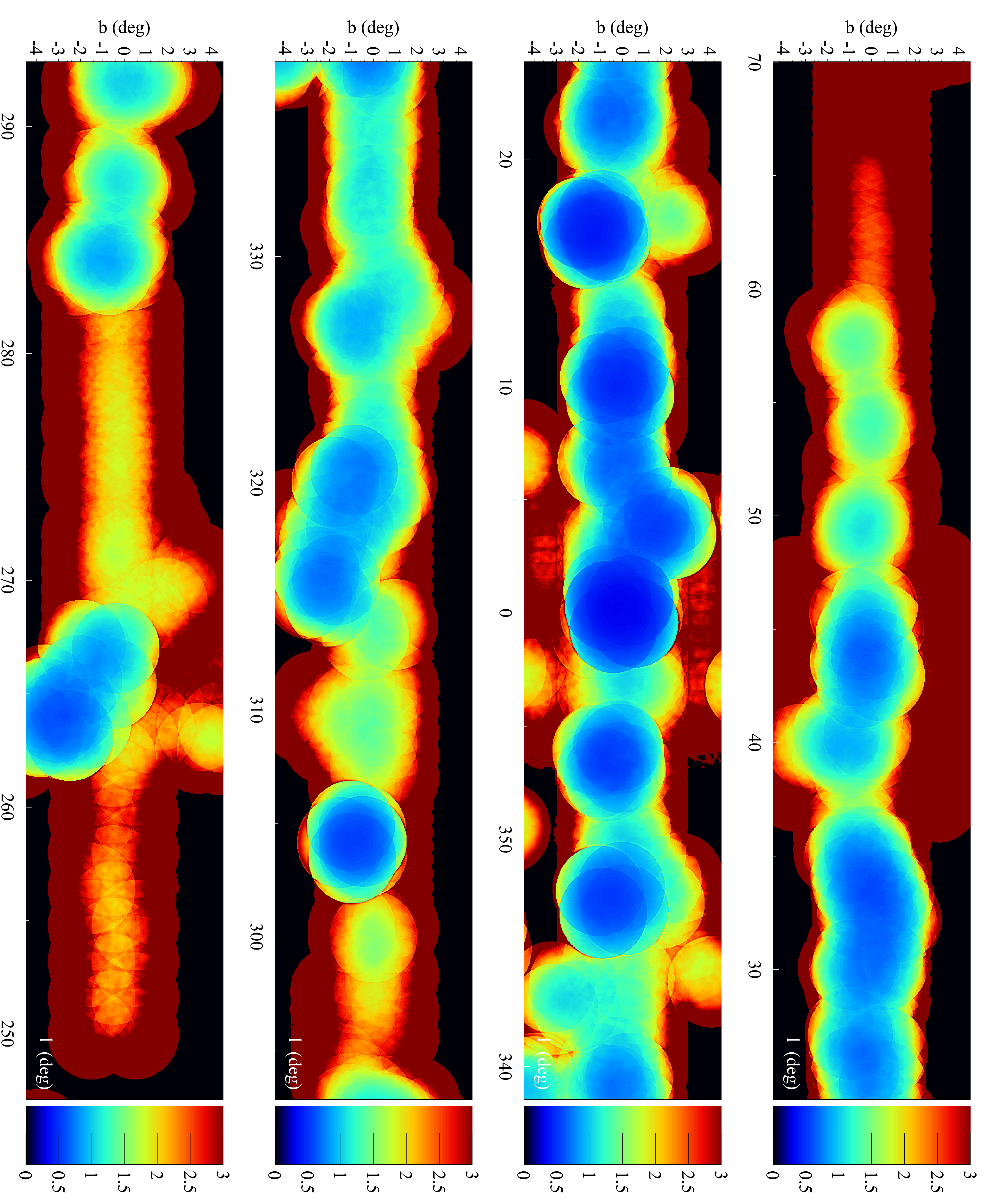}
\caption
{Sensitivity of H.E.S.S.~to point-like $\gamma$-ray sources
with an assumed spectral index of $2.3$, for a detection level of
$5\sigma$ pre-trial. The sensitivity is expressed as an integral flux
above 1 TeV in
units of $2.26\,\cdot\,10^{-9}\,\mathrm{m}^{-2}\,\mathrm{s}^{-1}$,
which amounts to $1\,\%$ of the Crab integral flux above 1 TeV.
}
\label{fig:sensitivitymap}
\end{figure}

Figure \ref{fig:sensitivitymap} depicts the current sensitivity to VHE
$\gamma$-ray sources, as an example
for point-like sources emitting a power-law spectrum with index 2.3 
and located at a Galactic latitude of b = $-$0.3$^\circ$,
the approximate average among known Galactic sources. The sensitivity
is below 2\% Crab for practically all of the longitude range l = 283$^\circ$  to
59$^\circ$  at this latitude. This can also be seen in a slice of this map, 
shown in Figure \ref{fig:sensitivityslice}.

\section{Source catalog}

Using the H.E.S.S. survey maps as input, we have implemented a
pipeline to generate 
a source catalog, which will be published alongside with the maps. The
aim is to have detection criteria as well as 
morphological and spectral analysis as uniform as possible, to make
this source catalog useful for the astronomical 
community to compare to data from other wavelengths,
e.g. Fermi~\footnote{http://fermi.gsfc.nasa.gov} or 
HAWK~\footnote{http://www.hawc-observatory.org}, for radio or 
X-ray data, as well as Galactic source population studies and diffuse emission
measurements. 

\begin{figure}
\includegraphics[width=\linewidth]{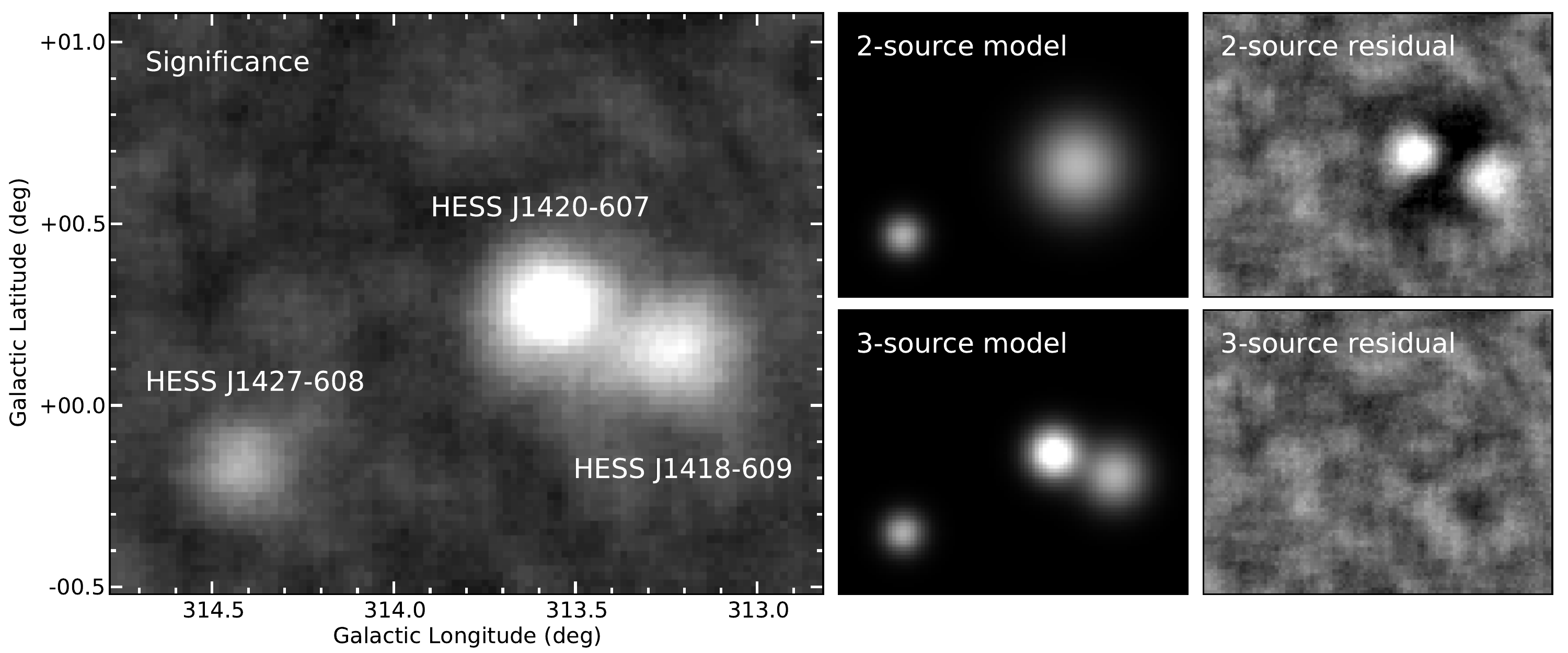}
\caption
{Iterative process of the source fitting in the {\it Kookaburra}
  region. Significance map (left) and model and residual images (small
  pictures on the right). It can be seen that the 3-source solution
  leaves a flat residual map and is therefore preferred. 
}
\label{fig:cat_method}
\end{figure}

To construct the catalog, we use a likelihood fit of the
H.E.S.S. counts map taking the exposure and 
point spread function as well as the estimated background into account.
To acknowledge the fact that most sources in the H.E.S.S. Galactic plane survey
region have been found to be extended, we are producing an extended
source catalog 
by modelling the excess as the superposition of symmetric Gauss-shaped
sources. The 
Gauss shape is not physically motivated, it was chosen as one commonly used
empirical shape in the absence of an expected source morphology
model. Figure~\ref{fig:cat_method} displays the fitting process. In
the left panel a significance map of the Kookaburra 
region and HESS~J1427$-$608 is shown, while the middle and right panel
show a 2 sources model and 3 sources fit model, respectively,
alongside with their resulting residual maps. The 
3 sources solution, i.e. the Kookaburra region splitting up into the
two sources HESS~J1420$-$607 and HESS~J1418$-$609, is significantly
better. Indeed it is believed that the emission  
from HESS~J1420$-$607 and HESS~J1418$-$609 is produced by two
different, large-offset PWNe, K3 powered by the high-spin down pulsar
PSR~J1420$-$6048 in case of HESS~J1420$-$607, and the Rabbit or R2, in
case of HESS~J1418$-$609~\cite{HESS_Kookaburra}. The third source,
HESS~J1427$-$607, is slightly extended and so far could not be
identified with a counterpart such as an SNR or PWN~\cite{HESS_Unid}.

For H.E.S.S. the background is dominated by the small fraction of
hadronic air showers that 
cannot be distinguished from gamma-ray-induced air showers.
After source detection and subsequent estimation of source
position and extension, for each catalog source a circular source
region containing most of the emission according to the best-fit
source model excess map is chosen, and the spectral analysis is
run independently of the maps. The
measurement of spectra is 
done by using the {\it reflected background method}~\cite{Berge2007}
(see right panel of 
Figure~\ref{fig:excl_bg}). 
In this method, multiple background regions (OFF regions, blue filled
circles) arranged in a circle are used for each trial  
source position (ON region, red filled circle), where each
OFF region has the same size and shape as the ON region and an
equal offset to the observation, or pointing, position. Due to the
equal offset of ON and OFF regions from the pointing direction of the system, no
radial acceptance correction is required with this method, making it
ideal for spectral analysis. Figure~\ref{fig:excl_bg} also illustrates
the motivation behind using two different sets of exclusion regions, as 
the {\it large} exclusion regions can be too large to allow for
spectral extraction in many cases, and the {\it small} exclusion regions are
used instead.

The difficult part about the catalog construction is the high source
density in the inner  Galaxy. There are several regions of
multi-degree-scale excess along the Galactic 
plane where it is not obvious how to best represent them in an extended
source catalog. Therefore we are investigating e.g. adding a criterion
to the catalog construction method that prevents large and bright
sources from decomposing into multiple, strongly overlapping
components.

\section{Galactic source population and recent discoveries}

\begin{figure}
\begin{centering}
\includegraphics[width=13cm]{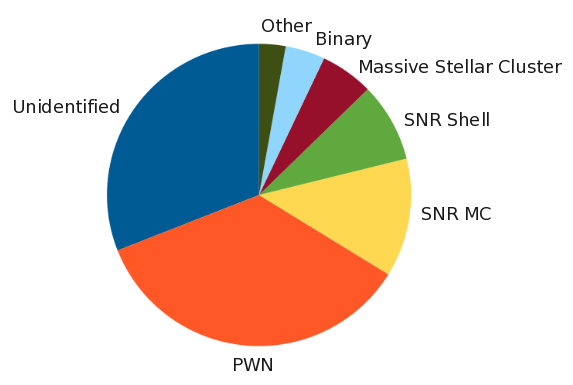}
\caption
{Galactic H.E.S.S. sources, status February 2013, classifications taken from
  TeVCat. Pulsar wind nebulae (PWN, orange) are the largest source class
  with $\sim$35\%, followed by supernova
  remnants, either interacting with a molecular cloud (SNR MC, yellow,
   $\sim$13\%)
  or exhibiting emission from their shell (SNR Shell, light
  green, $\sim$8\%). Besides a few massive stellar clusters (dark red,
  $\sim$6\%) and some  binary systems (light blue, $\sim$4\%) a large
  part of the H.E.S.S. source population remains unidentified (dark
  blue,  $\sim$31\%). 'Other' (dark green) comprises the
  globular cluster Terzan~5 and the high-frequency peaked BL Lac
  object HESS~J1943$+$213.
}
\label{fig:sources}
\end{centering}
\end{figure}

To date, 67 sources are listed in the catalog of published H.E.S.S. sources
\footnote{http://www.mpi-hd.mpg.de/hfm/HESS/pages/home/sources/}. 
Figure~\ref{fig:sources}  
shows a pie chart of the H.E.S.S. Galactic source population, status February
2013, where the source classification is taken from
TeVCat~\footnote{http://tevcat.uchicago.edu/}. While the largest
source class are pulsar wind nebulae (PWN, 
orange), followed by supernova remnants, either interacting with a
molecular cloud (SNR MC, yellow) or exhibiting emission from their
shell (SNR Shell, light green), there are to date only a few massive
stellar clusters (dark red) and binary systems (light blue) identified as
H.E.S.S. sources. A large part of the H.E.S.S. source population
remains ambiguous, therefore Unidentified (dark blue). It should be noted
that with further multi-wavelength data the distribution of this chart
will likely change, not only will sources migrate from being
unidentified to another source class, but in some cases possibly even
between the defined source classes. Therefore this chart represents
our knowledge at this point in time.

An example of a previously unidentified source that has now been
identified as a PWN is {\bf HESS~J1303$-$631}, which was the first
source classed as unidentified for H.E.S.S. Significant energy-dependent
morphology of this source, as well as the identification of an
associated X-ray PWN from XMM-Newton observations enable
identification of the VHE source as an evolved PWN associated with the
pulsar PSR J1301$-$6305~\cite{1303_follow}. 
 
{\bf W49B} is an SNR interacting with a molecular cloud, located in
the W49 region. W49B has one of the highest surface brightnesses in radio
of all the SNRs of this class in our Galaxy and is one of the
brightest ejecta-dominated SNRs 
in X-rays. Infrared observations evidenced that W49B is interacting
with molecular clouds and 
Fermi reported the detection of a coincident bright, high-energy
$\gamma$-ray source~\cite{Fermi_W49}. H.E.S.S. detected significant
emission from the W49 region, compatible with VHE emission from the
SNR W49B~\cite{HESS_W49}. The position of the emission is compatible
with the brightest part of the radio emission from the SNR as well as
with the GeV emission. Energy spectra in the GeV and TeV bands are in
very good agreement. Given the very high GeV luminosity, the GeV-TeV
connection, and the fact that the SNR is interacting with a dense
molecular cloud, a hadronic emission scenario is favored in the case
of W49B.

Finally, {\bf HESS~J1641$-$463} is an example for a source that
remains unidentified, despite of the existence of multi-wavelength
data~\cite{HESS_1641}. It is found within the
bounds of a radio SNR, however, the existing X-ray observations do not
provide additional support to this scenario due to the lack of
detection of an extended X-ray feature at the position of
HESS~J1641$-$463. In addition, the larger extension of
the SNR G338.5+0.1 as compared to the H.E.S.S. source, and the
relatively old age of the SNR inferred from its physical size suggests
that the emission might not be necessarily connected with the SNR but
rather with a PWN at its center, driven by an yet undetected
pulsar. Due to its small size, compatible with a point-like source for
H.E.S.S., the possibility that HESS~J1641$-$463 is a binary system
cannot be excluded.

\section{Outlook}
After nearly a decade of observing the Southern sky, H.E.S.S. is
ending its surveying 
program of the Galactic Plane. With the additional, much larger
fifth Cherenkov telesope in the centre of the H.E.S.S. array,
H.E.S.S. is entering phase II and will concentrate on deeper
observations with improved sensitivity and angular resolution. Here
we have shown the latest results of the HGPS and an overview over
the H.E.S.S. Galactic source population. We aim to present the entire data set
of the HGPS, in the form of maps and a catalog, and to make it accessible
to astronomers. With this large data set, population studies in the
VHE range become possible for the first time.

\section*{Acknowledgments}
The support of the Namibian authorities and of the University of
Namibia in facilitating the construction and operation of H.E.S.S. is
gratefully acknowledged, as is the support by the German Ministry for
Education and Research (BMBF), the Max Planck Society, the French
Ministry for Research, the CNRS- IN2P3 and the Astroparticle
Interdisciplinary Programme of the CNRS, the U.K. Science and
Technology Facilities Council (STFC), the IPNP of the Charles
University, the Polish Ministry of Science and Higher Education, the
South African Department of Science and Technology and National
Research Foundation, and by the University of Namibia. We appreciate
the excellent work of the technical support staff in Berlin, Durham,
Hamburg, Heidelberg, Palaiseau, Paris, Saclay, and in Namibia in the
construction and operation of the equipment.

\end{document}